# Photosynthesis re-wired on the pico-second timescale

Tomi K. Baikie[1+], Laura T. Wey[2+], Hitesh Medipally[3], Erwin Reisner[4], Marc M. Nowaczyk[3], Richard H. Friend[1], Christopher J. Howe[2*], Christoph Schnedermann[1*], Akshay Rao[1*], Jenny Z. Zhang[4*]

1 - Cavendish Laboratory, University of Cambridge, J. J. Thomson Avenue, Cambridge, CB3 0HE, UK
2 - Department of Biochemistry, University of Cambridge, Tennis Court Road, Cambridge, CB2 1QW, UK
3 - Plant Biochemistry, Ruhr-Universität Bochum, Universitätsstrasse 150, 44780 Bochum, Germany
4 - Department of Chemistry, University of Cambridge, Lensfield Road, Cambridge, CB1 2EW, UK

## Introductory Paragraph

Photosystems II and I (PSII and PSI) are the reaction centre-containing complexes that drive the light reactions of photosynthesis. PSII performs light-driven water oxidation (quantum efficiencies and catalysis rates of up to 80% and 1000 e$^-$ s$^{-1}$, respectively) and PSI further photo-energises the harvested electrons (quantum efficiencies of ~100%).[1,2] The impressive performance of the light-dependent components of photosynthesis has motivated extensive biological, artificial and biohybrid approaches to 're-wire' photosynthesis to enable higher efficiencies and new reaction pathways, such as H$_2$ evolution or alternative CO$_2$ fixation.[3–5] To date these approaches have focussed on charge extraction at the terminal electron quinones of PSII and terminal iron-sulfur clusters of PSI.[6–9] Ideally electron extraction would be possible immediately from the photoexcited reaction centres to enable the greatest thermodynamic gains. However, this was believed to be impossible because the reaction centres are buried around 4 nm within PSII and 5 nm within PSI from the cytoplasmic face.[10,11] Here, we demonstrate using *in vivo* ultrafast transient absorption (TA) spectroscopy that it is possible to extract electrons directly from photoexcited PSI and PSII, using both live cyanobacterial cells and isolated photosystems, with the exogenous electron mediator 2,6-dichloro-1,4-benzoquinone (DCBQ). We postulate that DCBQ can oxidise peripheral chlorophyll pigments participating in highly delocalised charge transfer (CT) states after initial photoexcitation. Our results open new avenues to study and re-wire photosynthesis for bioenergy and semi-artificial photosynthesis.

# Main

Ultrafast TA spectroscopy was carried out on suspensions of the cyanobacterium *Synechocystis* sp. PCC 6803 (hereafter *Synechocystis*) to study photosynthesis (**Figure 1A-C**). *Synechocystis* is a model organism for studying photosynthesis because of the endosymbiotic origin of the chloroplast in eukaryotic plants and algae.[12] Previously, it has been challenging to apply ultrafast TA spectroscopy on living plant or algal cells (typically 10 µm)[13] because of their scattering properties. *Nannochloropsis sp.* algal cells (2-5 µm)[14] have been demonstrated to lessen these scattering effects due to their small size.[15,16] Here, the use of *Synechocystis* cells (2 µm)[17] lessens these scattering effects further, and we were able resolve the photodynamics of photosynthesis.

Photoexcitation of *Synechocystis* cells with a 200 fs pump pulse centred at 450 nm revealed rich initial transient spectral features in the visible (550 – 800 nm). As shown in **Figure 1D**, we identified a prominent negative feature at 685 nm, which decayed within ~20 ps to 10% of its initial value. We attribute this feature to the ground-state bleaching of the photoactive reaction centres of PSI and PSII. Interestingly, a lower-energy feature at 715 nm grew within ~2 ps after photoexcitation and subsequently decayed on a longer picosecond timescale, indicating rapid excited state relaxation processes. Similarly, we observed higher-energy positive transient features (<680 nm), to which we also assign a contribution from the photo-induced absorption of phycobilisomes and carotenoids of the cells, as these features resemble those of *in vitro* steady state absorption measurements[18,19] and are absent from isolated photosystems (**Supplementary Figure 6**). We verified that the photosystems remained intact without photodamage, consistent over multiple biological replicates, and were in a closed state throughout these measurements (**Supplementary Figures 1-4**). We confirmed that the key spectral features arise from fully assembled photosystems (**SI Section 1**).

To probe the effect of exogenous electron mediators on the photodynamics of cyanobacterial cells, we repeated the same measurements with the addition of 2,6-dichloro-1,4-benzoquinone (DCBQ). Surprisingly, as shown in **Figure 1E**, we found that addition of DCBQ led to accelerated decay dynamics of the negative feature at 685 nm. Simultaneously, the signal rise observed at 715 nm was suppressed, with a noticeable effect evident within just 600 fs after photoexcitation. These observations demonstrate that DCBQ can alter the excited state decay pathways of the photosynthetic reaction centres in *Synechocystis* cells on a sub-ps timescale.

Previously, DCBQ has been thought to extract electrons only from the terminal electron acceptor site of PSII, the $Q_B$ pocket (**Figure 1C**). However, we note that the mid-point potential of DCBQ (+0.315 V vs SHE at pH 7)[20] makes it suitable to extract electrons from PSI also (P700*/P700$^+$ = -1.290 V, $F_B/F_B^-$ = -0.590 V)[21] and PSII (P680*/P680$^+$ = -0.660 V).[22] To rule out energy transfer from the reaction centres to DCBQ as the mechanism resulting in the observations shown in **Figure 1E**, we characterised the optical properties of suspensions of cells in the presence of DCBQ. As reported in **Figure 2B**, the absorption spectrum of cells includes prominent absorption bands at 450, 680 and 700 nm corresponding primarily to the chlorophylls in the cell's PSII and PSI complexes, while DCBQ absorbs at <300 nm. The lack of overlap in the absorption spectra of DCBQ and the intact cells indicates that energy transfer mechanisms in the form of Dexter or Förster resonances are not active (more detail **SI Section 2**).[16] DCBQ must hence act via an electron transfer mechanism.

It is known that all chlorophylls embedded within the reaction centres in PSII and PSI are energetically degenerate at room temperature. Photoexcitation thus results in a highly delocalised excited state shared across several chlorophylls.[23,24] Previous studies on isolated reaction centres from the photosynthetic purple non-sulfur bacterium *Rhodobacter sphaeroides* at 77 K showed that the initially excited state exhibits significant charge-transfer (CT) character and can form intermediate CT states within 200 fs.[25] Similarly, recent femtosecond crystallography results of the photosynthetic reaction centre of another purple non-sulfur bacterium *Blastochloris viridis* (formerly known as *Rhodopseudomonas viridis*) demonstrated electron transfer reactivity within 3 ps.[26]

In light of these observations, we postulate that photoexcitation of the chlorophyll pigments within the photosystems of *Synechocystis* at room temperature can also form highly delocalised intermediate CT states within the time-resolution of our measurement (200 fs), followed by electron transfer kinetics. Given that the cell absorption spectra for all DCBQ concentrations remained unchanged (**Supplementary Figure 19**) and only the $Q_B$ pocket in PSII is a docking site for DCBQ in the photosystems, we conclude that DCBQ does not tightly bind to the reaction centre chlorophylls. Instead, DCBQ most likely interacts with peripheral chlorophyll pigments protruding from the protein scaffold of the photosystems and interacting with the highly delocalised CT state formed after the initial photoexcitation (**Figure 2A**, also see **SI Section 3** for in-depth discussion).

Based on this assessment, we constructed a simple kinetic model (**Supplementary Figure 12**) and applied global analysis techniques to time-resolved TA data from cells to extract the relevant lifetimes. We then used this model to determine how efficiently DCBQ diverts electrons away from the native electron transfer chain (see **SI Section 1.4** for details). As highlighted in **Figure 2C**, our analysis yielded declining electron transfer lifetimes for increased DCBQ concentration. Based on our model and these values, we can estimate that 1 mM DCBQ diverts $17 \pm 6$ % of the photoexcited reaction centre population in wild-type *Synechocystis* cells. Critically, this electron transfer occurs as early as 600 fs **(Figure 1 D-E)**, which is consistent with our hypothesis that electrons can be extracted from rapidly formed, delocalised chlorophyll CT states.

Further support for this mechanistic picture stems from additional control experiments with 3-(3,4-dichlorophenyl)-1,1-dimethylurea (DCMU), which binds to the $Q_B$ pocket in PSII,[27] competing against DCBQ docking there and reduction thereafter. Here, the transient absorption measurements with DCBQ and DCMU yielded similar effects as with only DCBQ, indicating that DCBQ acts at the initial stages of the photosynthetic pathway, rather than only at the $Q_B$ site (**Figure 2D, SI Figure 13)**. This trend is also supported by analogous photoelectrochemistry measurements, where DCBQ was observed to extract electrons at a site that is alternative to the $Q_B$ site (approximately 30% of the photocurrent – see **SI Section 4**).

The ability of DCBQ to interact with the photosynthetic electron transport chain at longer timescales also was probed. Photoluminescence decay measurements revealed a reduction in luminescence lifetime with increasing DCBQ concentration, consistent with pulse amplitude modulation fluorometry studies of eukaryotic algal cells in the presence of benzoquinones (see **SI Section 5**).[8,20,28,29] This is in line with DCBQ interfering with the photosynthetic electron transport chain beyond the picosecond timescales explored in TA. Oxygen evolution measurements confirmed that PSII continues to perform oxidation of water *in vivo* in the presence of DCBQ at all concentrations tested in the TA experiments, suggesting the pathway for holes generated by photoexcitation remains active upon the addition of DCBQ (**SI Section 6**). However, a long-term cytotoxicity assay revealed that all concentrations of DCBQ greater than 200 µM are cytotoxic to the cells after 12 hours, but there was no correlation between the cytotoxicity of benzoquinones and their ability to re-direct electrons from the electron transport chain (**SI Section 7**).

We tested two other closely-related benzoquinones, phenyl-1,4-quinone (PPBQ) and 2,6-dimethyl-1,4-benzoquinone (DMBQ), which have comparable midpoint potentials to DCBQ for accepting electrons.[20,30] **Figure 2D** outlines that the early onset decay in the TA spectrum found for DCBQ was not observed in PPBQ- and DMBQ-treated cells (detailed analysis in **SI Section 1.5**). Similarly, complementary photoelectrochemistry experiments highlighted that DCBQ was the only benzoquinone screened that was reduced earlier than $Q_B$ in PSII *in vivo*, as it still mediated electron transfer to electrodes in the presence of DCMU (**SI Section 4**). The fact that PPBQ and DMBQ did not demonstrate early interactions with the reaction centres may originate from PPBQ and DMBQ exhibiting lower solubilities in aqueous solutions, causing these quinones to be sequestered in intracellular membranes and precipitating in aqueous compartments as they entered the cell,[28] or other poor association effects. These observations suggest that for an exogenous electron mediator to access the early photosynthetic electron transport chain for reduction, it must pass through the outer membrane and plasma membrane of the cell and also traverse the cytoplasm.

Finally, we set out to elucidate the action of DCBQ electron extraction from different stages of the photosynthetic electron transport chain. To this end, we carried out ultrafast TA spectroscopy on the photosystems *in vivo* with genetically engineered cells lacking either PSI or PSII intact protein complexes (**Figure 3A**). These measurements were further complemented *in vitro* with isolated photosystems, with and without DCBQ. A full analysis is presented in **SI Section 1.6**.

While the mutant cells and isolated photosystems allow us to deconvolve the spectral features of the reaction centres, the changes in the lifetimes highlight the importance of observing the photodynamics of wild-type cells to obtain accurate insights into the photosynthetic electron transport chain. Following photoexcitation at 450 nm, the PSI-less mutant cells revealed spectral features markedly different from wild-type cells (**Figure 3B** – top panel). The PSI-less cells exhibited a long-lived positive signal (less absorption) at its absorption maximum at 680 nm, lasting longer than a nanosecond. This extended ground state bleach likely arises as the natural charge extraction pathways were blocked, or not fully functional, due to the absence of functional PSI. In contrast, the PSII-less cells exhibited more similar dynamics to wild-type cells, with the familiar low-energy spectral feature >690 nm (**Figure 3B**).

We then studied the effect of DCBQ addition on these mutant cells (full details **SI Section 1.6**). Upon the addition of 1 mM DCBQ to the PSI-less cells, we found a reduction from 74 ps to 53 ps in the lifetime of the short component of the signal (**Figure 3C** – top panel). This suggests that electron extraction by DCBQ treatment from PSII in PSI-less cells occurs over a much longer timescale than observed in the wild-type cells. Upon the addition of DCBQ to the PSII-less cells, we found a reduction in the lifetimes which closely resembled the behaviour in wild-type cells.

Measurements of living cells are advantageous compared to the those of *in vitro* systems due to the robustness of the cells and the ability to study altered pathways produced via genetic engineering. However, *in vitro* measurements are useful in identifying spectral features due to their relative simplicity. *In vitro* measurements of PSI and PSII revealed similar spectral features (**SI Section 1.6**) and lifetime reductions compared to mutant cells upon the addition of DCBQ, further supporting our assignments and proposed mechanism. Considering the *in vitro* and *in vivo* measurements together, we conclude that in the experimental arrangement described here, we primarily resolve electron transfer from PSI to DCBQ. This is in line with the more accessible nature of PSI compared to PSII. Although PSII is also weakly accessible to DCBQ in the PSI-less cells, no corresponding long-lived signal is observed in the wild-type cells that can be directly assigned to PSII with sufficient statistical certainty.

## Conclusion and outlook

Taken together, our results show that *in vivo* electron transfer from photosynthesis to exogenous electron mediators, here demonstrated using DCBQ, is possible directly from the initially photoexcited states of PSI and PSII, i.e., from the earliest possible step in the photosynthetic electron transport chain. This opens new avenues for the rational design of strategies to re-wire photosynthesis, for example to couple directly such photoexcited states to $H_2$ evolution or $CO_2$ reduction via a more reducing mediator. Moreover, such strategies are not limited to exogenous mediators. Endogenous mediator strategies could be engineered, as well as other bio-hybrid approaches. These results also call for a re-examination of mediated electron transfer and its mechanisms, for example in protein-film and microbial electrochemical studies, where DCBQ is commonly employed as an exogenous electron mediator.[31–35] Finally, our work highlights that *in vivo* ultrafast spectroscopic measurements are feasible and can reveal rich information on photoexcited dynamics of living systems.

# Figures

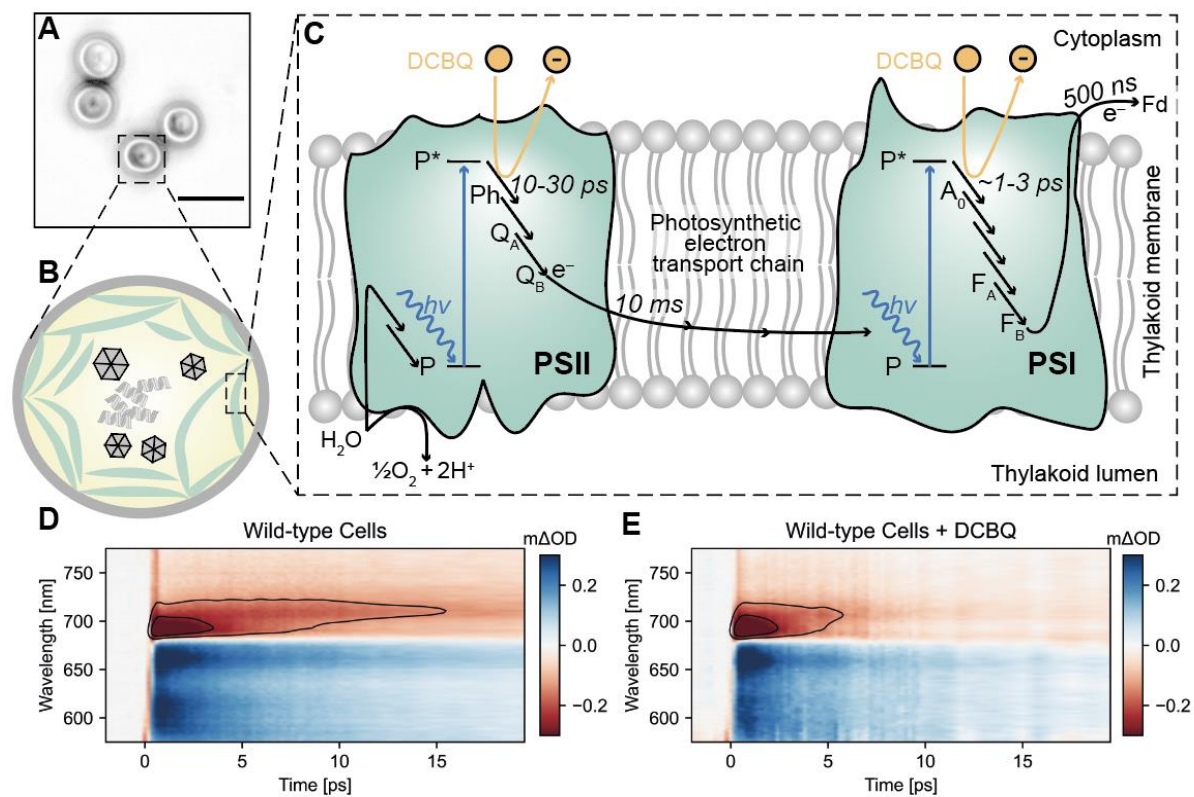

**Figure 1: Exogenous electron mediator acts on the picosecond time scale in living cells.**
**A –** Bright field microscopy image of *Synechocystis sp.* PCC6803 cells. Scale bar is 5 μm. **B –** Schematic representation of a cell showing intracellular thylakoid membranes (green). **C –** Schematic representation of a simplified photosynthetic electron transport chain. The chain starts with photosystem II (PSII), in which a special pair of chlorophylls collectively known as P680 is photoexcited to P680* directly or by energy transfer from other chlorophylls. An initial charge separated state with the primary electron acceptor pheophytin (Ph) [P680$^{+\cdot}$–Ph$^{-\cdot}$] is formed that has a lifetime of 10-30 ps,[36] which drives water oxidation. The extracted electrons exit PSII via the terminal plastoquinones associated with the PsbD and PsbA subunits: $Q_A$ and $Q_B$. The electrons are shuttled along the electron transport chain where the first diffusion step takes approximately 10 ms.[37] The electrons are received by photosystem I (PSI) and with further input of light, the P700 is photoexcited to P700* and an initial charge separated state with the primary electron acceptor chlorophyll $A_0$ [P700$^{+\cdot}$–$A_0^{-\cdot}$] is formed that has a lifetime of approximately 1-3 ps in the A-branch.[38] The electrons exit PSI via the terminal iron-sulfur clusters bound to the PsaC subunit: $F_A$ and $F_B$. The electrons are shuttled via ferredoxin (Fd) which takes approximately 500 ns,[39] and then onto various pathways, including the Calvin Benson cycle. Upon the addition of the electron mediator 2,6-dichloro-1,4-benzoquinone (DCBQ, orange circle) a new electron transfer pathway forms (orange arrow). Ultrafast transient absorption spectroscopy map of **D -** wild-type cells and **E –** cells with DCBQ (5 mM) between -2 and 20 ps excited at 450 nm. As a guide for the eye, two contours have been drawn on each map at the same signal intensity. These highlight the reduced lifetime of the 685 nm feature and suppression of the 715 nm rise upon the addition of DCBQ. Maps are of one sample, and the change in dynamics upon addition of DCBQ is representative of multiple biological replicates (wild-type cells, n = 8, wild-type cells + DCBQ, n = 5).

.

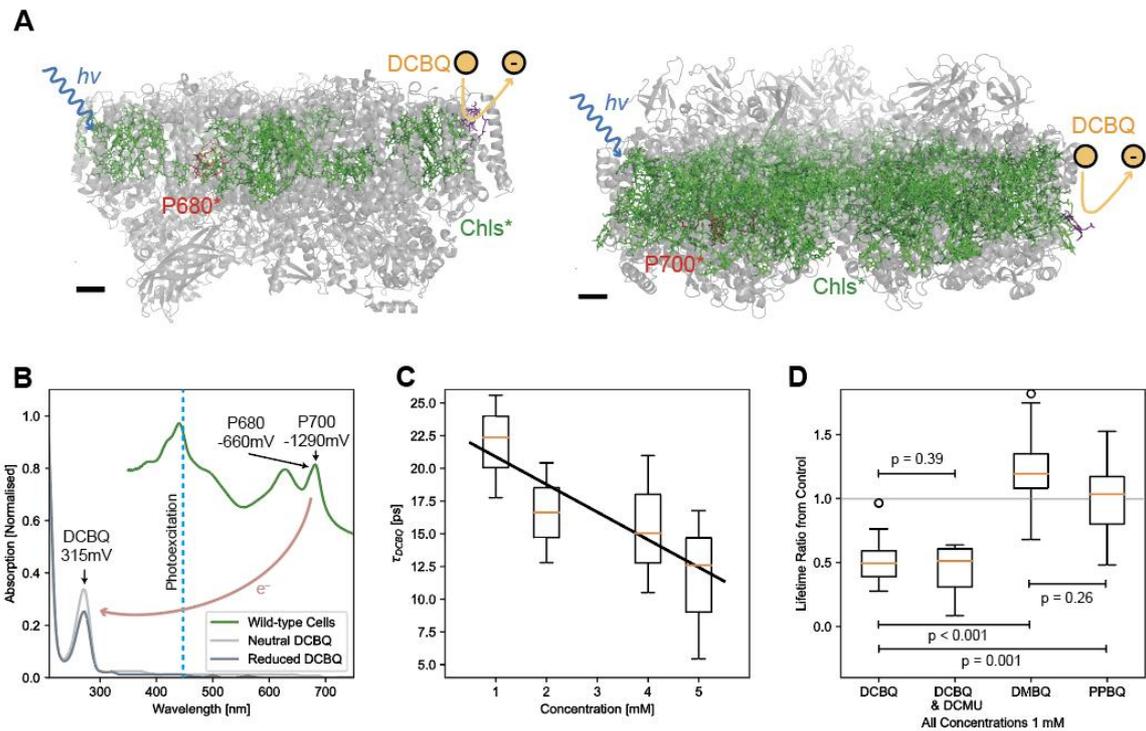

**Figure 2: Action of quinone electron mediators on wild-type cells.**
**A –** Schematic representation of the proposed mechanism of early DCBQ reduction by photosystem II (PSII) and photosystem I (PSI). The chlorophylls are photoexcited (Chls*, green), including the special pairs of chlorophylls (P680* and P700*, red in the left most monomer), in a delocalised excited state. The initial charge separated states are formed and rapidly delocalised across the chlorophylls. DCBQ (orange circle) is reduced via a periphery chlorophyll (purple) that protrudes from the protein scaffold into the thylakoid membrane. PSII crystal structure by Young *et al.* (2016) (Protein Data Bank ID: 5TIS);[40] PSI crystal structure by Jordan *et al.* (2001) (Protein Data Bank ID: 1JB0).[10] Scale bar 10 Å. **B –** Absorption spectra of *Synechocystis* cells (green), and reduced and neutral DCBQ (greys). **C –** Retrieved electron transfer lifetimes to DCBQ. Detail of the model in SI Section 2.4.2. The black line is a linear fit to the medians. **D –** Lifetime change upon the addition of DCBQ, DMBQ, PPBQ, and DCMU with DCBQ. Detail of model in SI Section 1.5. Presented are the ratios of the additive lifetimes over from samples where no quinone was added (control). A ratio of 1 (grey line) indicates no effect. Student t-test comparisons are overlaid. In C and D, the box extends from the lower to upper quartile values of the data, the whiskers extend from the box to show the range of the biological replicates (n = 5, except DCBQ & DCMU n = 2), and the orange line represents the median.

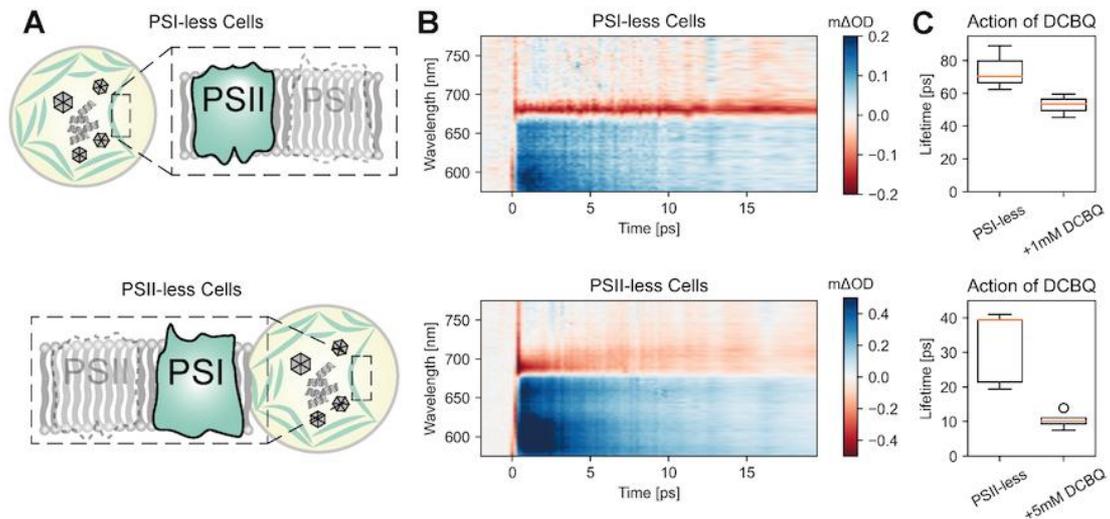

**Figure 3: Action of DCBQ on wild-type cells genetically modified to have only one type of photosystem.**
**A –** Schematic representation of photosystem-less (PS-less) mutants analysed (top – PSI-less, bottom – PSII-less). **B** – TA spectra between -2 and 20 ps of cell mutants excited at 450 nm under same conditions as wild-type cells in their TA spectrum in Figure 1D (top – PSI-less, bottom – PSII-less). The PSII-less spectrum (bottom) closely resembles that of the wild-type cells. Maps are of one sample, representative of multiple biological replicates (PSI-less: n = 4, PSII-less: n = 5). **C** – Lifetimes determined from global analysis in the PS-less cells and with the addition of DCBQ (top – PSI-less, bottom – PSII-less). The box extends from the lower to upper quartile values of the data, the whiskers extend from the box to show the range of the biological replicates (PSI-less + DCBQ: n = 4, PSII-less + DCBQ: n = 3), and the orange line represents the median.

## Data Availability

The data underlying all figures in the main text are publicly available from the University of Cambridge repository at [URL added in proof]

## Code Availability

All code used in this work is available from the corresponding authors upon reasonable request.

## Acknowledgements

We acknowledge Professor Wim Vermaas (Arizona State University, USA) for the gift of the photosystem-less mutants used in this study; and Professor William Rutherford (Imperial College London) for the gift of isolated PSII as well as valuable discussions on this project. We acknowledge Professor Frederic Lemaitre (ENS, France) and Professor Peter Rich (UCL, UK) for helpful discussions about exogenous benzoquinones and photosynthetic microorganisms. C.S. and T.K.B. thank Dr. Victor Gray for insightful discussion at the start of the project. T.K.B. gives thanks to the Centre for Doctoral Training in New and Sustainable Photovoltaics for financial support L.T.W. acknowledges financial support from the Cambridge Trust. C.S. acknowledges financial support by the Royal Commission of the Exhibition of 1851. We acknowledge financial support by the BBSRC (BB/R011923/1 to J.Z.Z.). We acknowledge financial support from the EPSRC and the Winton Program for the Physics of Sustainability, as well as from the Deutsche Forschungsgemeinschaft (DFG) within the framework of the Research Training Group 2341 'MiCon'. This project has received funding from the European Research Council (ERC) under the European Union's Horizon 2020 research and innovation programme (grant agreement no. 758826 and no. 764920).

## Author Contributions

T.K.B. and L.T.W. contributed equally to the work and came up with the concept of examining cyanobacteria using the techniques described. C.S. and AR supervised the spectroscopy, C.J.H. supervised the cell work, J.Z.Z came up with the research question and managed the project. T.K.B. performed the TA and TCSPC experiments and the analysis, including devising the kinetic model and preparing the figures. L.T.W. chose and prepared the samples for TA and TCSPC, performed the photoelectrochemistry, oxygen evolution, cytotoxicity and microscopy experiments, and did protein crystal structure analysis. H.M. prepared isolated PSI. A.R., C.J.H., J.Z.Z. and M.M.N. provided funding and resources. T.K.B., L.T.W., C.S. and J.Z.Z. wrote the initial draft. All authors contributed to analysis, review and editing.

## Materials and Methods

### Biological Samples
#### Cyanobacterial cells
#### Cell Culture and Growth Conditions

*Synechocystis sp.* PCC 6803 (*Synechocystis*) cells were cultured photoautotrophically under 50 µmol$_{photons}$ m$^{-2}$ s$^{-1}$ of continuous white light at 30°C in BG11 medium supplemented with 10 mM NaHCO$_3$.[41] Liquid cultures were bubbled with air and shaken at 120 rpm. 1.5% (w/v) agar was used in solid medium. Culture growth was measured by attenuance at 750 nm (OD$_{750}$). Culture chlorophyll concentration (nmol Chl *a* ml$^{-1}$) was calculated from absorbances at 680 nm and 750 nm: (A$_{680}$-A$_{750}$) × 10.814.[42] All measurements were taken using a UV-1800 Spectrophotometer (Shimadzu).

## Photosystem-less Mutants

The *Synechocystis* photosystem-less mutants were a gift from Prof. Wim Vermaas (Arizona State University, USA). The modifications to the culturing protocol for the strains are outlined in **Materials and Methods Table 1**.

**Materials and Methods Table 1 – Photosystem-less (PS) mutants used in this study**

| Name | Gene(s) knocked out | Antibiotics | Culture conditions | Reference |
|---|---|---|---|---|
| **Wild-type** | - | - | BG11 no additives, 30˚C, 40 µmol$_{photons}$ m$^{-2}$ s$^{-1}$ light | - |
| **PSI-less** | *psaAB*- | 10 µg/mL chloramphenicol | BG11 plus 5 mM glucose, 30˚C, 5 µmol$_{photons}$ m$^{-2}$ s$^{-1}$ light | [43] |
| **PSII-less** | *psbDIC*-/*psbDII*- | 10 µg/mL chloramphenicol, 10 µg/mL spectinomycin | BG11 plus 5 mM glucose, 30˚C, 40 µmol$_{photons}$ m$^{-2}$ s$^{-1}$ light | [44] |

## Olive mutant

The *Synechocystis* Olive mutant was generated previously by markerless disruption of the *cpcBAC1C2* genes.[45] As reported previously the Olive mutant has no phycocyanin (PC) discs in the phycobilisome (PBS). The Olive mutant still has the allophycocyanin (AC) core of the PBS. Deletion of the entire PBS results in extremely poor growth,[45] so this mutant was not analysed in this study.

## Isolated Photosystems

In this study, whole cells of cyanobacterium *Synechocystis* were compared against photosystem II (PSII) and photosystem I (PSI) protein complexes isolated from the cyanobacterium *Thermosynechococcus elongatus*. PSII isolated from mesophilic *Synechocystis* is significantly more unstable than those extracted from *T. elongatus*. Previous studies have compared the photoelectrochemistry of PSII *in vitro* versus *in vivo* from *T. elongatus* and *Synechocystis*, respectively, for similar stability reasons.[34] The purified *T. elongatus* PSII was provided by the group of Prof. William A. Rutherford (Imperial College London, UK). PSII dimers were isolated from *T. elongatus* using methods previously reported.[46] The PSII stock had a chlorophyll concentration of 2.4 mg/mL in 2-(N-morpholino)ethanesulfonic acid (MES) storage buffer (10% glycerol, 15 mM MgCl$_2$, 15 mM CaCl$_2$, 1 M betaine, 40 mM MES (pH 6.5)). Purified PSI trimers were isolated from *T. elongatus* using methods previously reported.[47] The PSI stock had a chlorophyll concentration of 9.07 mg/mL in MES storage buffer. Isolated photosystems were stored in small aliquots in a liquid N$_2$ Dewar flask, and a new aliquot was thawed on each day of experiments. Photoelectrochemistry and spectroscopy measurements on isolated photosystems were conducted in MES buffer (50 mM KCl, 15 mM CaCl$_2$, 15 mM MgCl$_2$, 40 mM MES (pH 6.5)). The *T. elongatus* was cultured at 60˚C, the spectroscopy suite was maintained at less than 21˚C and the photoelectrochemistry set-up was controlled at 25˚C. Preparation methods for spectroscopy has been reported previously.[48]

## Chemicals

Unless otherwise stated, all materials used throughout this research were purchased from commercial suppliers and used without further purification. The exogenous electron mediators used in this study were (**Materials and Methods Figure 1**): 2,6-dichloro-1,4-benzoquinone (DCBQ, 98%, Sigma-Aldrich), phenyl-1,4-quinone (PPBQ, 95%, ACROS Organics) and 2,6-

dimethyl-p-benzoquinone (DMBQ, 97% Sigma-Aldrich). Benzoquinone candidates were selected from previous studies with *Chlamydomonas reinhardtii* by the Group of Prof. Frédéric Lemaître.[8,20,28,29] The photosynthetic electron transport chain (PETC) inhibitor used in this study was 3-(3,4-dichloro-phenyl)-1,1-dimethylurea (DCMU, 98%, Alfa Aesar). DCMU was used as an inhibitor of PSII at the $Q_B$ site.[27] Stocks of 100 mM DCBQ and 100 mM DCMU were made in DMSO for addition to biological samples. Stocks of 10 mM DMBQ and PPBQ were made in ethanol for addition to biological samples, with 8.2 mM SDS where specified.

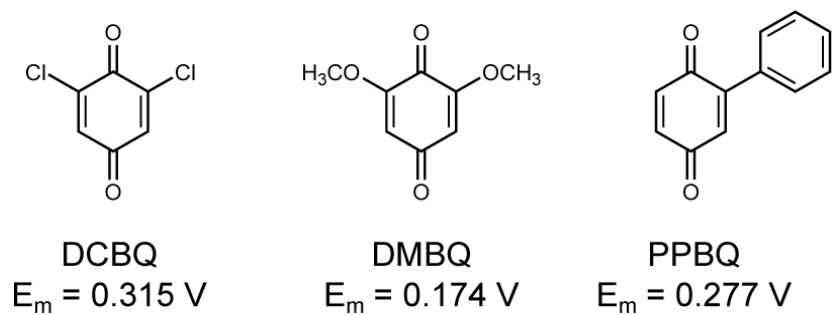

**Materials and Methods Figure 1** - *Chemical structures and mid-point potentials of electron mediators used in this study* Mediators: 2,6-dichloro-1,4-benzoquinone (DCBQ), phenyl-1,4-quinone (PPBQ), 2,6-dimethyl-p-benzoquinone (DMBQ). Mid-point potentials ($E_m$) from the literature versus SHE and at pH 7.[20,30]

## Ultrafast transient absorption spectroscopy

Picosecond transient absorbance measurements were performed as previously reported.[49,50] In brief, we used an Yb-based amplified system (PHAROS, Light Conversion) providing 14.5 W at 1030 nm and a 38 kHz repetition rate. The probe beam is generated by focusing a portion of the fundamental in a 4 mm YAG substrate to generate a white light which spans from 520 to 900 nm. The pump beam is generated by seeding a portion of the fundamental to a narrow band optical parametric oscillator (ORPHEUS-LYRA, Light Conversion). The pump pulse was set to 450 nm, unless otherwise indicated. The sample solutions were placed in 1 mm path length cuvettes (Hellma). The pump and probe beams were focused to a size of $280 \times 240\ \mu m$ and $55 \times 67\ \mu m$, respectively. The probe is delayed relative to the pump using a computer-controlled translation stage (Newport), and a chopper wheel (Thorlabs) modulated the pump beam to gain access to differential transient absorption spectra. The probe pulses transmitted through the sample were detected on a single-shot basis by a line camera (Stemmer Imaging). All measurements presented in our work were acquired within 15 min of loading the sample. Additional control experiments exposed the cells to the laser for 120 min (8-fold longer).

## Time-correlated single-photon counting

To record the time-resolved photoluminescence decay of the samples, time-correlated single-photon counting (TCSPC) was performed. Samples were excited with a pulsed laser (PicoQuant LDH-400-B and LDH-470-B, (at up to 40 MHz, typically operated at 2.5 MHz)) of at either 407 nm or 470 nm, with the resulting photoluminescence decay collected on at $680 \pm 10$ nm. Diode lasers were controlled by a trigger box/power supply unit (PDL 800-B, PicoQuant). TCSPC utilised an emission spectrometer (Lifespec-ps unit, VTC900 PCI card, Edinburgh Instruments) with a multi-channel plate detector (R3809U-50, Hamamatsu). The instrument response was determined by scattering excitation light into the detector using a piece of frosted glass; a value of 265 ps was obtained. Pulse energy was 2 pJ. Pulse width of the diode lasers were measured to be 80 ps (full width at half maximum, FWHM).

## Spectroelectrochemistry

Solutions of DCBQ (1 mM) in BG11 medium and MES buffer were prepared. The DCBQ solution (1 mL) was loaded into an optically transparent thin-layer electrochemical cuvette with a pseudo reference electrode (Ag wire). Cyclic voltammetry was performed (applied potential 0.5 V to -0.2 V vs pseudo reference Ag wire, scan rate: 0.05 V s$^{-1}$) to determine that the working electrode poised at -0.1 V would reduce the DCBQ. The UV-Vis spectrometer was baselined with a solution of 1% DMSO (v/v) in BG11 medium or MES buffer as appropriate. The spectrum of the DCBQ solution was measured without applying a potential (neutral form). Chronoamperometry experiments were performed at -0.1 V. At 30 min into the chronoamperometry the spectrum of the DCBQ (doubly reduced form) solution was measured.

## Photoelectrochemistry

All photoelectrochemical measurements were performed at 25°C under atmospheric conditions using an Ivium Technologies CompactStat, with an Ag/AgCl (saturated KCl) reference electrode (corrected by + 0.197 V for SHE) and a platinum mesh counter electrode. Chronoamperometry experiments were performed at an applied potential 0.5 V vs SHE. Chronoamperometry experiments were performed at a sampling rate 1 s$^{-1}$, under light/dark cycles using a collimated LED light source (50 µmol$_{photons}$ m$^{-2}$ s$^{-1}$, approximately 1 mW cm$^{-2}$ equivalent, 680 nm, Thorlabs).

### Protein-film Photoelectrochemistry

Inverse opal indium-tin oxide (IO-ITO) electrodes with 750 nm pore sizes were prepared using a previously reported method.[46] Protein-film photoelectrochemistry was performed as previously reported.[34] In brief, a one-in-three dilution of the 2.4 mg/mL (77 µM) stock solution of isolated PSII was made immediately before adsorption on the electrodes to give a final concentration of 25.6 µM. A small aliquot (1 µl) of the new solution was drop-cast onto 750 nm IO-ITO electrodes and left to stand for 15 min in a closed petri dish in the dark before being used in photoelectrochemical experiments. The electrolyte was MES buffer (pH 6.5). Light/dark cycles (15 s on/15 s off) were used.

### Cell photoelectrochemistry

IO-ITO electrodes with 10 µm macropores and 3 µm interconnecting channels at a thickness of 40 µm were prepared using the method previously reported.[34] Photoelectrochemistry of cyanobacterial cells was performed as previously reported.[51] In brief, planktonic cultures of early stationary phase cyanobacterial cells at attenuance at 750 nm (OD$_{750}$) of ca. 1 were concentrated by centrifugation at 5000 g for 10 min, the supernatant removed, and the pellet resuspended in fresh BG11 medium to a concentration of 150 nmol Chl *a* ml$^{-1}$. This solution (250 µl) was dropcast onto the IO-ITO electrodes and left overnight at room temperature in a humid chamber in the dark to allow cell penetration and adhesion, yielding cell-loaded electrodes that were used for analysis 16 h later.  The electrolyte was BG11 medium (pH 8.5). Light/dark cycles (60 s on/90 s off) were used.

## Oxygen evolution measurements

Oxygen evolution measurements were measured using a Clark electrode consisting of an Oxygraph Plus Electrode Control Unit, S1 Oxygen Electrode Disc, DW2/2 Electrode Chamber and a LED1 High Intensity LED Light Source (Hansatech Instruments). Measurements were performed on 1.5 ml samples of wild-type *Synechocystis* cells containing 10 µg ml$^{-1}$ of Chl *a* in BG11 supplemented with different concentrations of DCBQ.  Measurements were collected at 25°C with 1 min darkness, followed by 1 min of 1500 µmol$_{photons}$ m$^{-2}$ s$^{-1}$ light at 627 nm (equivalent to 28.65 mW cm$^{-2}$). The rate of oxygen production in the dark was subtracted from that in the light and normalised to Chl *a* content. Data was collected from five biological replicates, each with three technical replicates.

### Cytotoxicity assays

Wild-type *Synechocystis* cells (5 nmol Chl *a*) were incubated with different concentrations of exogenous mediator for 24 h. *Synechocystis* cells incubated in BG11 with no electron mediator or in BG11 with 5 % (v/v) DMSO solvent under the same conditions were used as controls. Following incubation, the cells were resuspended in fresh BG11 and their concentration was standardised to an optical density at 750 nm ($OD_{750}$) of 0.5. Aliquots (10 µl) of three serial dilutions (x 1, x $10^{-3}$, x $10^{-6}$) were spotted on BG11 agar plates, which were then incubated for 1 week at 30˚C and 50 µmol$_{photons}$ m$^{-2}$ s$^{-1}$ light. The growth of the cells pre-incubated with mediator was compared to the controls to assess the cytotoxicity of the mediators.

### Statistics

The ratios of the lifetime of samples upon the addition of benzoquinones and DCMU compared to where nothing was added were compared using student t-tests.

## Supplementary Information

The supplementary Information contains further information on UV/Vis, transient absorption and TCSPC experiments of biological controls and their replicates, an overview over the data analysis used for these methods, more detailed structural studies as well as spectroelectrochemical results.

## Author information

Cavendish Laboratory, University of Cambridge, J. J. Thomson Avenue, Cambridge, CB3 0HE, UK
Tomi K. Baikie, Richard H. Friend, Christoph Schnedermann, Akshay Rao

Department of Biochemistry, University of Cambridge, Tennis Court Road, Cambridge, CB2 1QW, UK
Laura T. Wey, Christopher J. Howe

Laura T. Wey
Present address: University of Turku, Turku, Finland

Department of Chemistry, University of Cambridge, Lensfield Road, Cambridge, CB1 2EW, UK
Erwin Reisner, Jenny Z. Zhang

Plant Biochemistry, Ruhr-Universität Bochum, Universitätsstrasse 150, 44780 Bochum, Germany
Hitesh Medipally, Marc M. Nowaczyk

## Corresponding Authors

Correspondence to:
Christopher Howe – ch26@cam.ac.uk
Christoph Schnedermann – cs2002@cam.ac.uk
Akshay Rao – ar525@cam.ac.uk
Jenny Zhang – jz366@cam.ac.uk

## Competing Interests

The authors declare no competing interests.